\documentclass[aps,pre,floatfix,showpacs,amsmath,amssymb,twocolumn,superscriptaddress]{revtex4-1}
\usepackage{graphicx}

\begin{document}
\title{Glass-Transition Properties from Hard Spheres to Charged Point 
Particles}

\author{Anoosheh Yazdi}
\affiliation{Institut f\"ur Materialphysik im Weltraum,
Deutsches Zentrum f\"ur Luft- und Raumfahrt, 51170 K\"oln, Germany}
\author{Alexei Ivlev}
\affiliation{Max-Planck-Institut f{\"u}r extraterrestrische Physik, 
85741 Garching, Germany}
\author{Sergei Khrapak}
\affiliation{Max-Planck-Institut f{\"u}r extraterrestrische Physik, 
85741 Garching, Germany}
\author{Hubertus Thomas}
\affiliation{Max-Planck-Institut f{\"u}r extraterrestrische Physik, 
85741 Garching, Germany}
\author{Gregor E. Morfill}
\affiliation{Max-Planck-Institut f{\"u}r extraterrestrische Physik, 
85741 Garching, Germany}
\author{Hartmut L\"owen}
\affiliation{Institut f{\"u}r Theoretische Physik II: Weiche Materie, 
Heinrich-Heine-Universit{\"a}t D{\"u}sseldorf, Universit{\"a}tsstrasse 1, 
40225 D{\"u}sseldorf, Germany}
\author{Adam Wysocki}
\affiliation{Institut f{\"u}r Theoretische Physik II: Weiche Materie, 
Heinrich-Heine-Universit{\"a}t D{\"u}sseldorf, Universit{\"a}tsstrasse 1, 
40225 D{\"u}sseldorf, Germany}
\author{Matthias Sperl}
\affiliation{Institut f\"ur Materialphysik im Weltraum,
Deutsches Zentrum f\"ur Luft- und Raumfahrt, 51170 K\"oln, Germany}

\date{\today}
\begin{abstract}

The glass transition is investigated in three dimensions for single and 
double Yukawa potentials for the full range of control parameters. For 
vanishing screening parameter, the limit of the one-component plasma is 
obtained; for large screening parameters and high coupling strengths, the 
glass-transition properties crossover to the hard-sphere system. Between 
the two limits, the entire transition diagram can be described by 
analytical functions. Different from other potentials, the 
glass-transition and melting lines for Yukawa potentials are found to 
follow shifted but otherwise identical curves in control-parameter space.

\end{abstract}


\maketitle

\section{Introduction}

The physics of ordered (crystals) and disordered (glasses) solid states 
and their interrelation has been subject to investigations in various 
model systems \cite{Pusey1986,Ivlev2012}. Such model systems capture 
typical features of more complex matter and often allow for the variation 
of the interparticle interactions to explore physical regimes otherwise 
not accessible. A qualitatively strong variation concerns the 
distinction between hard and soft repulsion as in the Yukawa potential 
which describes the range from excluded-volume to charge-based 
interactions.

Yukawa potentials are realized in both colloidal suspensions 
\cite{Bitzer1994,Beck1999,Heinen2011} and complex plasmas 
\cite{Ivlev2012}, and since in complex plasmas the damping can be tuned, 
this offers a way for the comparison of Brownian and Newtonian dynamics 
with the same particle-particle interaction in experimental systems 
\cite{Gleim1998}. While in sterically stabilized colloidal suspensions, 
the interaction can typically be well-approximated by the hard-sphere 
interaction \cite{Pusey1986}, for charged particles in suspensions, 
hard-sphere plus Yukawa interaction is more appropriate. In complex 
plasmas, the average interparticle distance compared to the particles' 
diameters is typically large enough to allow for an approximation of 
point-like particles and hence a screened Coulomb potential for point 
particles is appropriate. In addition to the screening length, in complex 
plasmas also a second repulsive length scale arises from the 
non-equilibrium ionization-recombination balance 
\cite{Khrapak2010,Wysocki2010} which gives rise to a double Yukawa 
potential at interparticle distances $r$ as
\begin{equation}\label{eq:pot} 
\frac{U(r)}{k_\text{B}T} = \frac{\Gamma}{r} \left[ \exp(-\kappa r) 
	+ \epsilon \exp(-\alpha\kappa r) \right]\,.
\end{equation} 
Distance $r$ is given in units of the mean interparticle distance 
$1/\sqrt[3]{\rho}$ with the density $\rho = N/V$ for $N$ particles in a 
volume $V$. The coupling parameter is $\Gamma = Q^2 
\sqrt[3]{\rho}/(k_\text{B}T)$, with the charge $Q$, and $\kappa = 
1/(\lambda\sqrt[3]{\rho})$ is the inverse of the screening length 
$\lambda$. The second (longer-ranged) Yukawa potential is specified by a 
relative strength $\epsilon$, and a relative inverse screening length 
$\alpha < 1$. In the limit of vanishing screening, one recovers the 
one-component plasma (OCP), the simplest model that exhibits 
characteristics of charged systems \cite{Hansen1986}. Motivated by the 
success of mode-coupling theory for ideal glass transitions (MCT) for the 
hard-sphere system (HSS), cf. \cite{Sperl2005}, in the following, the 
glass-transition shall be calculated within MCT \cite{Goetze2009}. Since 
for time-reversible evolution operators, i.e., Newtonian and Brownian 
dynamics, the glassy dynamics within MCT are identical \cite{Szamel1991}, 
the calculations are applicable to both complex plasma and charged 
colloids.

\section{Methods}

We consider a system of $N$ point-like particles in a volume $V$ of 
density $\rho = N/V$ interacting via the pairwise repulsive potential in 
Eq.~(\ref{eq:pot}). We investigate the glass transitions in two cases: the 
single Yukawa ($\epsilon=0$) and the double Yukawa potential ($\epsilon > 
0$). Within MCT, the glass transition is defined as a singularity of the 
form factor $f_q = \lim_{t \rightarrow \infty} \phi_q(t)$ that is the 
long-time limit of the density autocorrelation function. In the liquid 
state, $f_q$ is zero, while in the glass state, $f_q > 0$. At the 
transition, the form factors adopt their critical values $f_q^c \geq 0$. 
$f_q$ is the solution of \cite{Bengtzelius1984}
\begin{equation}\label{eq:fq} 
\frac{f_q}{1-f_q} = {\cal F}_q[f_k]\,, 
\end{equation} 
which is the long time limit of the full MCT equations of motion. $f_q$ is 
distinguished from other solutions of the Eq.~(\ref{eq:fq}) by its maximum 
property, thus it can be calculated using the iteration 
$f^{(n+1)}_q/(1-f^{(n+1)}_q) = {\cal F}_q[f^{(n)}_k]$ \cite{Franosch1997} 
with $f^{(0)}_k=1$ and the memory kernel given by
\begin {equation}\label{eq:Fq}
{\cal F}_q[f_k]=  \frac{1}{16 \pi^3} \int d^3k \frac{S_q S_k S_p}{q^4}
[\mathbf{q}\cdot\mathbf{k} c_k+\mathbf{q}\cdot\mathbf{p}c_p]^2 f_k f_p\,,
\end {equation}
where $\mathbf{p}=\mathbf{q}-\mathbf{k}$; all wave vectors are expressed 
in normalized units. Note that the number density does not appear 
explicitly in the kernel ${\cal F}_q$, since we express the length scales 
in units of $1/\sqrt[3]{\rho}$. 

The only inputs to the Eq.~(\ref{eq:Fq}) are the static structure factors 
$S_q$. The Fourier transformed direct correlation functions $c_q$, are 
related to structure factors through the Ornstein-Zernike (OZ) relation 
\begin{equation}\label{eq:oz} 
\gamma_q=\frac{c_q^2}{1-c_q}, 
\end{equation} 
where the spatial Fourier transform of $\gamma_q$ is $\gamma(r) = h(r) - 
c(r)$ and $h(r)$ is the total correlation function which is related to 
structure factor through $S_q=1+h_q$. We close the equations by the 
hypernetted-chain (HNC) approximation,
\begin{equation}\label{eq:HNC} 
c(r)=\exp{[-U(r)/(k_BT)+\gamma(r)]}-\gamma(r)-1\,, 
\end{equation}
 where $U(r)$ is the interaction potential. It was found earlier that HNC 
captures well various structural features for repulsive potentials, 
especially also for the OCP \cite{Ng1974}. For the HSS, the quality of HNC 
is known to be inferior to the Percus-Yevick (PY) approximation in certain 
thermodynamic aspects \cite{Hansen1986}, so we expect HNC to vary in 
performance for different parameter regions of the Yukawa potentials in 
Eq.~(\ref{eq:pot}).

We solve Eq.~(\ref{eq:oz}) and Eq.~(\ref{eq:HNC}) by iteration and use the 
usual mixing method in order to ensure convergence \cite{Hansen1986}. We 
iterate $n$ times from an initial guess, $c^{(0)}(r)$, until a 
self-consistent result is achieved, i.e.,
\begin{equation}
\left[\int_0^R|c^{(n+1)}(r)-c^{(n)}(r)|^2\,\mathrm{d}r\right]^{1/2} < 
\delta,
\end{equation}
with $\delta = 10^{-5}$,
where $R$ is the cut-off length of $c(r)$. We employ $R=47.1239$ and a 
mesh of size $M=2396$ points. Consequently, the resolution in real and 
Fourier space is $\Delta r=R/M=0.0197$ and $\Delta q=\pi/R=0.0667$, 
respectively. We use an orthogonality-preserving algorithm for the 
numerical calculation of Fourier transforms \cite{Lado1971}. For a 
particular $\kappa$ we begin the computation of $c(r)$ at a small coupling 
parameter $\Gamma$, successively increase $\Gamma$, and use the outcome as 
an initial guess for the subsequent calculation.

\section{Single-Yukawa Potential}

\subsection{Glass-Transition Diagram}

\begin{figure}[hbt]
\includegraphics[width=\columnwidth]{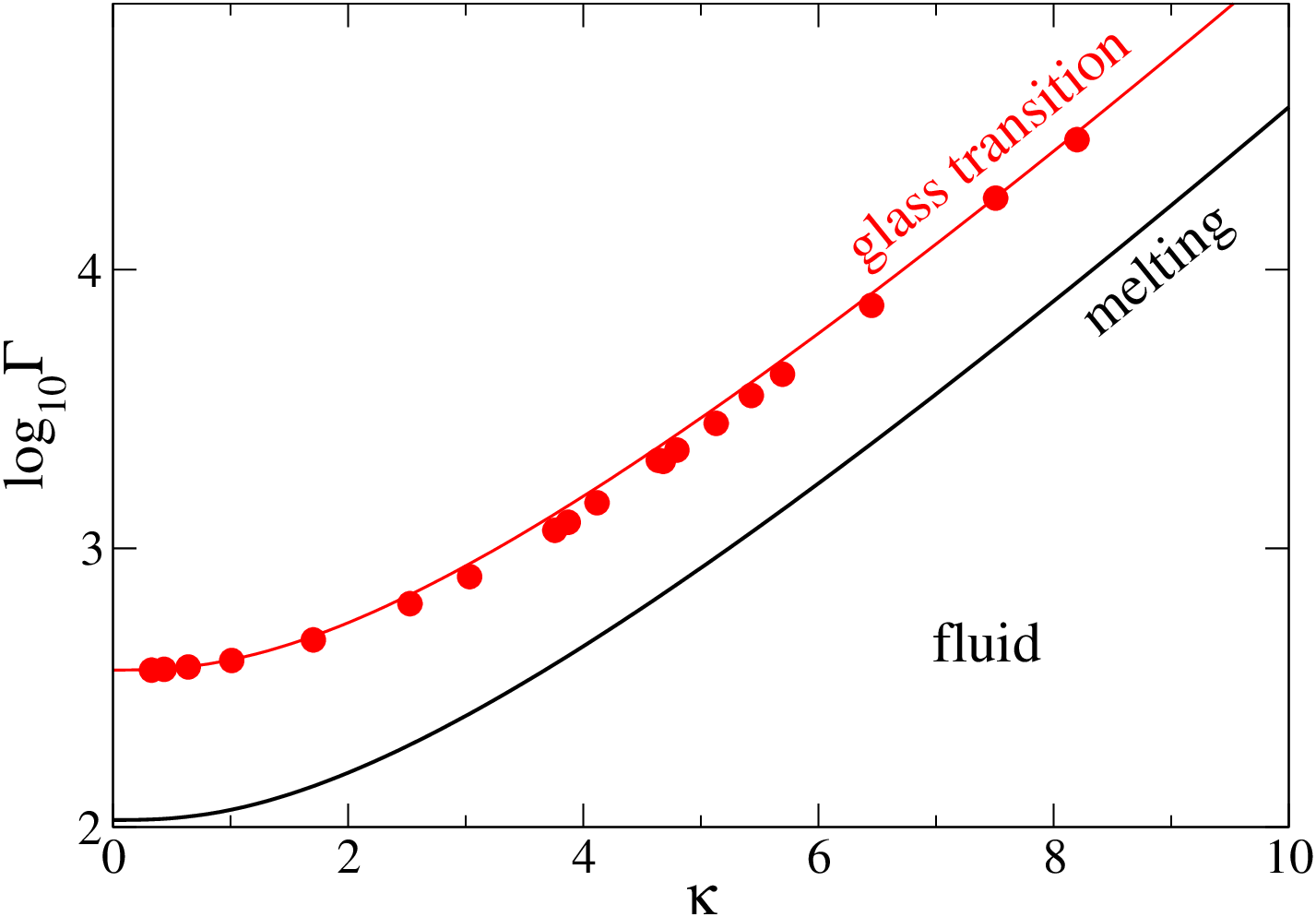}
\caption{\label{fig:PD1Y}Glass-transition diagram for the single Yukawa 
potential (filled circles). Transition points are shown together with the
full curve exhibiting Eq.~(\ref{eq:scale}). For comparison, a similar 
curve is shown for the melting of the crystal.}
\end{figure}

The MCT results for the single Yukawa case are shown in 
Fig.~\ref{fig:PD1Y}. The filled circles for different $\Gamma$ and 
$\kappa$ indicate the glass transition points calculated by 
Eq.~(\ref{eq:fq}). For $\kappa\rightarrow 0$, the glass transition for the 
OCP limit is found at $\Gamma^c_\text{OCP}=366$. When screening is 
introduced for $\kappa > 0$, the glass-transition line moves to higher 
critical coupling strengths $\Gamma^c(\kappa)$. Figure~\ref{fig:PD1Y} 
shows for reference the melting curve for weakly screened Yukawa systems, 
described by $\Gamma(\kappa) = 106\,e^{\kappa}/\left(1 + \kappa + 
\kappa^2/2\right)$ \cite{Vaulina2000,Vaulina2002}. This expression has 
been suggested originally on the basis of the Lindemann-type arguments, 
cf. \cite{Lindemann1910}. The Lindemann criterion states that the 
liquid-crystal phase transition occurs when in the crystal the 
root-mean-square displacement $\langle \delta r^2\rangle$ of particles 
from their equilibrium positions reaches a certain fraction of the mean 
interparticle distance. Within the simplest one-dimensional harmonic 
approximation this yields the scaling $U''(r=1)\langle \delta 
r^2\rangle/T\simeq {\rm const.}$, where primes denote the second 
derivative with respect to distance. Applied to the Yukawa interaction 
this leads to the melting curve above, where the value of the constant is 
determined from the condition $\Gamma\simeq 106$ at melting of the OCP 
system ($\kappa=0$) \footnote{Note that $\Gamma\simeq 172$ if the 
Wigner-Seitz radius $a = \sqrt[3]{3/4\pi\rho}$ is used as a unit length 
instead of $1/\sqrt[3]{\rho}$.}. This expression for the melting curve is 
widely used due to its particular simplicity and reasonable accuracy: 
Deviations from numerical simulation data of Ref.~\cite{Hamaguchi1997} do 
not exceed several percent, as long as $\kappa\lesssim 8$. Moreover, 
similar arguments can be used to reasonably describe freezing of other 
simple systems, e.g. Lennard-Jones-type fluids~\cite{Khrapak2011}. 
Remarkably, when comparing the predicted glass-transition with the melting 
curve, one observes that both transition lines run in parallel. The 
glass-transition line is described by the function
\begin{equation}\label{eq:scale}
\Gamma^c(\kappa) = \Gamma^c_\text{OCP}\,e^{\kappa}
	\left(1+\kappa+\kappa^2/2\right)^{-1}\,,
\end{equation}
which is shown as solid line in Fig.~\ref{fig:PD1Y}, i.e., the glass 
transition is found at 3.45 of the coupling strength of the melting curve.

The fit quality given by Eq.~(\ref{eq:scale}) is remarkable for two 
distinct reasons: First, the potential changes quite drastically along the 
line from a long-ranged interaction at low $\kappa$ to the paradigmatic 
hard-sphere system at very large $\kappa$ to be detailed below. Such 
simplicity along control-parameter dependent glass-transition lines is not 
to be expected and not observed for other potentials, cf. the square-well 
system \cite{Dawson2001,Sperl2004}. Second, the non-trivial changes along 
the transition lines are apparently quite similar for the transition into 
ordered and disordered solids alike, and Eq.~(\ref{eq:scale}) applies to 
both. For the mentioned square-well system, ordered and disordered solids 
have no such correlation \cite{Sperl2004}.

Since both MCT and the structural input involve approximations, typically 
the glass transitions are found for higher couplings than predicted, the 
deviation is around 10\% in the densities for the HSS \cite{Goetze2009}. 
While one can expect that absolute values for transition points need to be 
shifted to match experimental values \cite{Sperl2005}, the qualitative 
evolution of glass-transition lines with control parameters is usually 
quite accurate and even counterintuitive phenomena like melting by cooling 
have been predicted successfully \cite{Dawson2001}. Hence, we assume the 
description of the liquid-glass transitions in the single Yukawa system to 
be qualitatively correct.

\subsection{Glass-Form Factors}

\begin{figure}[hbt]
\includegraphics[width=\columnwidth]{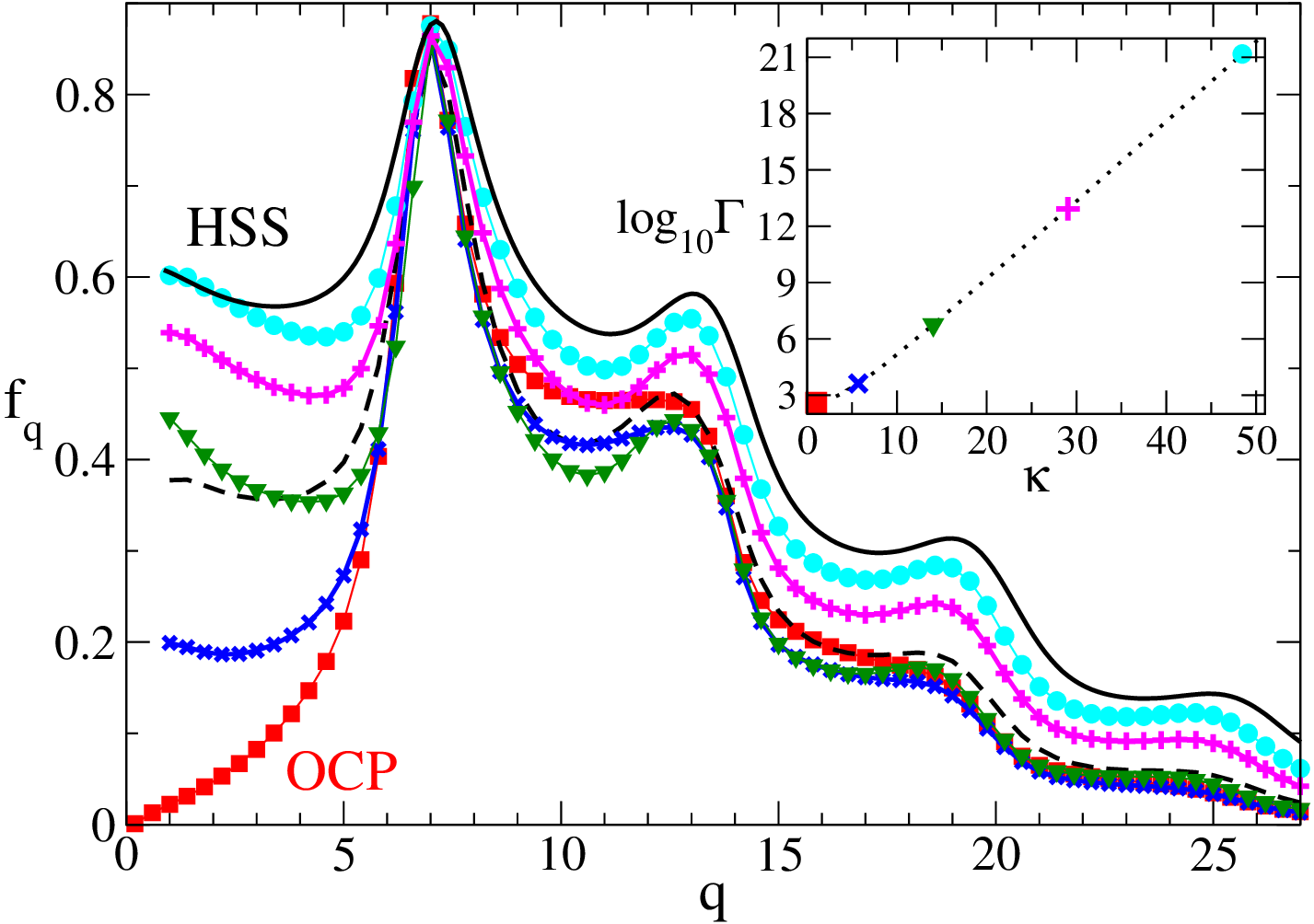}
\caption{\label{fig:fq1Y}Critical glass-form factors $f_q$ for the glass 
transition in the single Yukawa system. For increasing screening 
parameter $\kappa$, the inset shows the location of the respective 
transition points on the MCT-transition line, cf. Fig.~\ref{fig:PD1Y}, 
with the same symbols as in the main panel. The full curve shows the 
solution for the HSS within the HNC approximation. The result 
for HSS within the PY approximation \cite{Franosch1997} is shown dashed.}
\end{figure}

The different points on the glass-transition lines shall be discussed in 
detail in the following. For the well-known case of the glass transition 
in the HSS, the critical form factors are shown by a full curve in 
Fig.~\ref{fig:fq1Y}. Different from earlier results calculated for $S_q$ 
within the PY approximation \cite{Franosch1997}, here we also show the HSS 
within the HNC approximation to be consistent with the Yukawa results. The 
control parameter for the HSS is the packing fraction $\varphi = \rho
(\pi/6)\,d^3$ with the hard-core diameter $d$ as the unit of length. For 
HNC, the transition point is found at a packing fraction of 
$\varphi^c_\text{HSS} = 0.525$. This value as well as the behavior of 
$f_q$ in Fig.~\ref{fig:fq1Y} is very close in HNC and the PY approximation 
where $\varphi^c_\text{HSS} = 0.516$ \cite{Franosch1997}. It is seen that 
the distribution of $f_q$ is dominated by a peak at interparticle 
distances which indicate the cage effect \cite{Goetze2009,Franosch1997}; 
oscillations for higher wave vectors follow this length scale in a way 
similar to the static structure factor. For both PY and HNC, the peak 
positions for $f_q$ coincide, for the principal peak even the peak heights 
are almost identical. For HNC, the $f_q$ are typically above the PY 
solutions resulting in a 10\% larger half-width of the distribution of the 
$f_q$ for the HNC. The predicted deviations between HNC and PY are mostly 
indistinguishable when comparing to experiments except for the small-$q$ 
limit where experimental results favor the PY-MCT calculation, cf. 
\cite{Megen1995,Goetze2009}.

For the Yukawa potential, overall the critical form factors exhibit 
similar features as for the HSS. Different from the HSS, in the OCP limit 
the form factors vanish for the limit $q\rightarrow 0$. This anomaly for 
charged systems corresponds to the small wave-vector behavior in the 
static structure $S_q\propto q^2$ for $q\rightarrow 0$ \cite{Hansen1986}. 
Since in the OCP, mass and charge fluctuations are proportional to each 
other, the conservation of momentum implies the conservation of the 
microscopic electric current, and hence no damping of charge fluctuations 
in the long wave-length limit. Considering Eq.~(\ref{eq:Fq}) we shall 
demonstrate, that $f_q \propto q^2$ for small wave vectors.

Denoting $\theta$ as the angle between $\mathbf{q}$ and $\mathbf{k}$ we
can expand the direct correlation functions as:
\begin{equation}\label{eq:expanc}
c_{|\mathbf{q}-\mathbf{k}|}= c_k - c'_kq \;\text{cos}\; \theta+
\frac{1}{2} q^2 \text{cos}^2\theta c''_k- \frac{1}{6} q^3
\text{cos}^3\theta c'''_k
\end{equation}
where the primes represent the respective first, second and third 
derivatives of $c_k$ with respect to $k$. Substituting 
Eq.~(\ref{eq:expanc}) into Eq.~(\ref{eq:Fq}) leads to
\begin{subequations}\label{eq:Fqexp}
\begin {equation}\label{eq:Fqlimitmore}
{\cal F}_q[f_k] \rightarrow S_q \alpha +q^2 S_q \beta\,,
\end {equation}
where \cite{Bayer2007}
\begin{equation}
\alpha=\frac{1}{4\pi^2} \int_0^\infty dk k^2 S_k^2 [c_k^2 +
\frac{2}{3} k c_k c'_k+\frac{1}{5} k^2 {c'_k}^2] f_k^2 \,,
\end{equation}
and 
\begin{equation}\label{eq:beta}
\begin{split}
\beta=&\frac{1}{4\pi^2} \int_0^\infty dk k^2 
S_k^2 [\frac{1}{3}
{c'_k}^2+\frac{1}{28} k^2 {c''_k}^2+\frac{2}{5} k c'_k c''_k
\\&+\frac{1}{3}
c'_k c''_k+\frac{1}{15} k c_k c'''_k+\frac{1}{21} k^2 c'_k c''_k] 
f_k^2\,.\end{split}
\end{equation}
The term linear in $q$ in Eq.~(\ref{eq:Fqlimitmore}) vanishes. 
\end{subequations}
Similarly, 
the small-$q$ expansion of the static structure factor in the OCP reads 
\cite{Baus1980}
\begin{equation}\label{eq:sqlimit}
 S(q)= \frac{q^2}{k_D^2}+\frac{q^4}{k_D^4}[c^R(0)-1]+{\cal O}(q^6)
\end{equation}
where $k_D^2= 4 \pi \Gamma$ represents the inverse Debye length, and 
$c^R(q)=c(q)-c^S(q)$ is the regular term of the direct correlation 
function, assuming that at large distances particles can only be weakly 
coupled, which creates the singular term $c^S(q)=-U(q)/k_\text{B}T$. From 
Eq.~(\ref{eq:Fqlimitmore}) and Eq.~(\ref{eq:sqlimit}) we get 
\begin {equation}\label{eq:Fqlimitmoreocp}
{\cal F}_q= q^2 \frac{\alpha}{k_D^2} + q^4[\frac{\beta}{k_D^2}
        +\frac{\alpha}{k_D^4}(c^R(0)-1)]+{\cal O}(q^6)\,.
\end{equation}
From Eq.~(\ref{eq:fq}) one can conclude that $f_q$ has the same limit as 
$F_q$, hence we have shown that $f_q\propto q^2$ for vanishing $q$.

For non-vanishing screening, $\kappa > 0$, the small-$q$ behavior of the 
form factors is characterized by finite intercepts at $q=0$. This regular 
behavior is ensured by the $q\rightarrow0$ limit of $c^S_q = 
-4\pi\Gamma/(q^2 + \kappa^2)$. For larger wave vectors, the $f_q$ first 
decrease in comparison to OCP -- cf. $\kappa = 5.7$ ($\times$) and 14.0 
($\blacktriangledown$) in Fig.~\ref{fig:fq1Y} -- before increasing beyond 
the OCP result for $\kappa \gtrsim 30$. For very large screening, the form 
factors of the Yukawa potential apparently approach the HSS case.

\subsection{HSS Limit}

\begin{figure}[htb]
\includegraphics[width=\columnwidth]{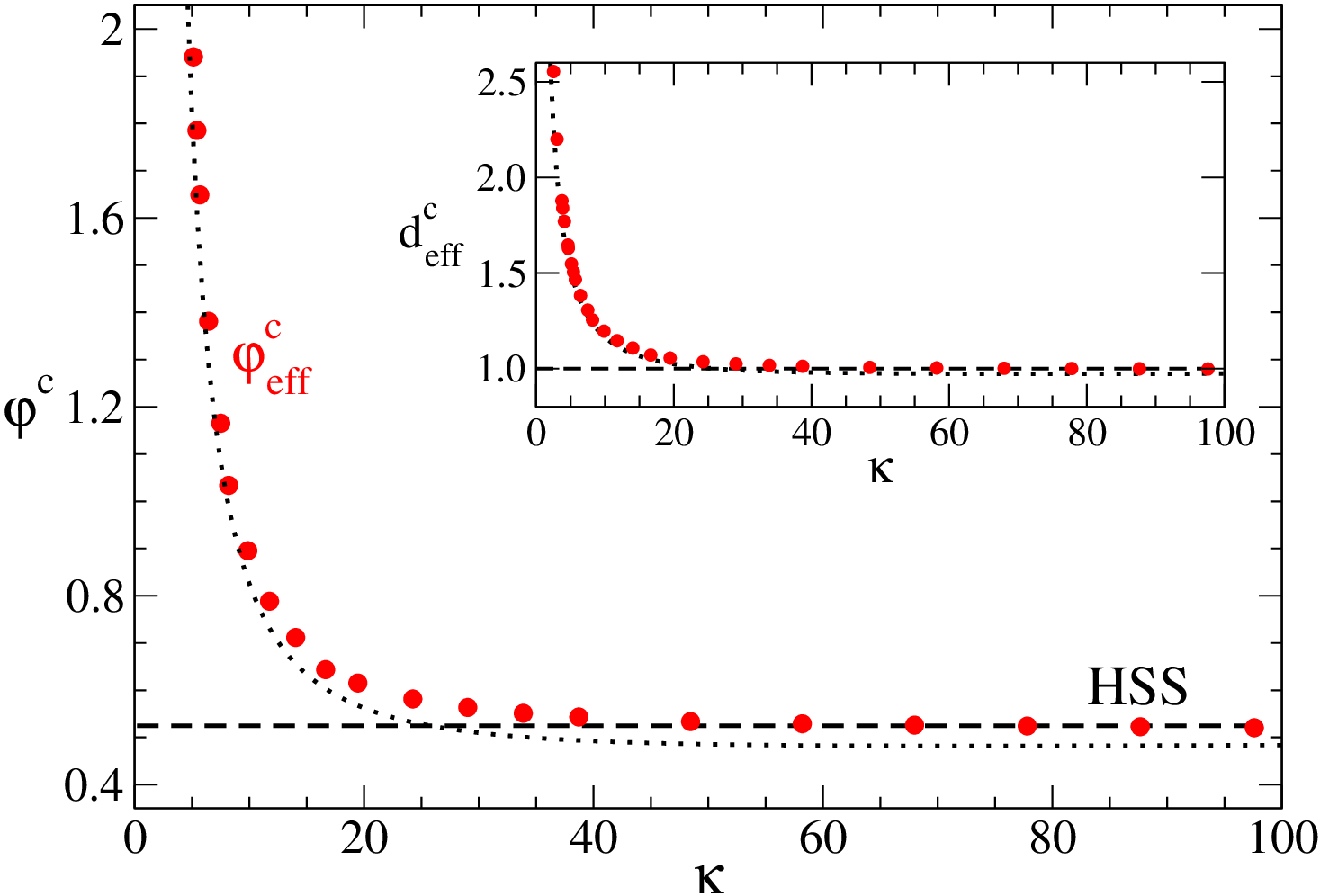}
\caption{\label{fig:phieff}Effective packing fraction 
$\varphi^c_\text{eff}$ for Yukawa potentials along the transition line in 
Fig.~\ref{fig:PD1Y}. The horizontal dashed line shows the HSS-HNC limit of 
$\varphi^c_\text{HSS} = 0.525$. The inset shows the effective hard-sphere 
diameter, $d^c_\text{eff} = \ln\Gamma^c/\kappa$ equivalent to the 
effective densities. In both plots, the dotted curves display the 
small-$\kappa$ asymptotes derived from Eq.~(\ref{eq:scale}). }
\end{figure}

By setting $U(d_\text{eff})/k_\text{B}T\sim 1$ for $\epsilon = 0$ in 
Eq.~(\ref{eq:pot}) one can define an effective diameter that becomes a 
well-defined hard-core diameter for $\kappa\rightarrow\infty$. Along the 
glass-transition line $\Gamma^c(\kappa)$ the effective packing fraction 
and diameter are given (with logarithmic accuracy) by
\begin{equation}\label{eq:HSSlimit}
\varphi^c_\text{eff} = 
\frac{\pi}{6}\left(\frac{\ln\Gamma^c}{\kappa}\right)^3\,,\quad
d^c_\text{eff} = \ln\Gamma^c/\kappa\,,
\end{equation}
where only the definition of the packing fraction has been used. 
Figure~\ref{fig:phieff} displays the effective packing fractions along the 
single-Yukawa transition line up to $\kappa\approx 100$. For small 
$\kappa$, the large effective diameter yields considerable 
overlaps among the particles and hence a packing fraction beyond unity. 
The effective hard-sphere diameter can be seen in the inset of 
Fig.~\ref{fig:phieff}. For $\kappa\gtrsim 40$ the Yukawa potentials'
effective diameter $d^c_\text{eff}$ reaches its asymptotic value.
Together with the findings on the $f_q$ this establishes the crossover of 
the glass-transition properties of the Yukawa system to the hard-sphere 
limit. The relation in Eq.~(\ref{eq:scale}) fits effective diameters and 
densities well for smaller $\kappa\lesssim 10$ and underestimates the 
calculated values for larger $\kappa$, as expected.

\section{Double-Yukawa Potential}

\subsection{Glass-Transition Diagrams}

\begin{figure}[bht]
\includegraphics[width=\columnwidth]{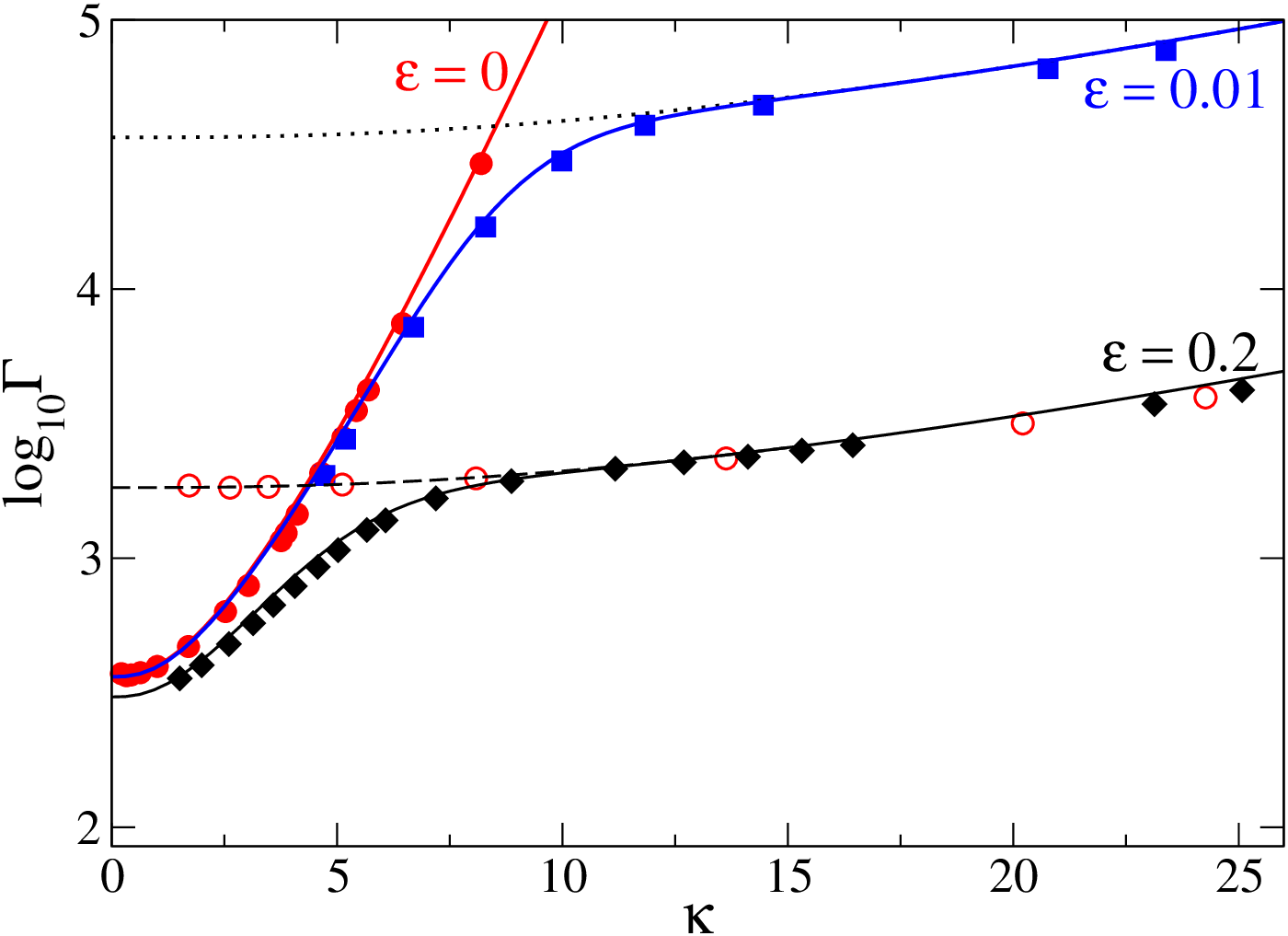} 
\caption{\label{fig:PD2Ys}Glass transition diagram for double Yukawa 
potentials with $\alpha = 0.125$, $\epsilon = 0.2$ (diamonds) and 0.01 
(squares). The single Yukawa data (filled circles) is shown together with 
the analytical description by Eq.~(\ref{eq:scale}) (solid curve labeled 
$\epsilon = 0$). The single Yukawa points are scaled according to 
Eq.~(\ref{eq:scale2Y}) for $\epsilon = 0.2$, and shown by open circles. 
Dotted  and dashed curves represent scaled versions of 
Eq.~(\ref{eq:scale}) for $\epsilon = 0.01$ and $\epsilon = 0.2$, 
respectively. The solid curves labeled $\epsilon = 0.01$ and $\epsilon = 
0.2$, respectively, show the solution of Eq.~(\ref{eq:scaleall}). 
} 
\end{figure}

Progressing towards the double Yukawa potentials, we show in 
Fig.~\ref{fig:PD2Ys} the results of MCT calculations for the same relative 
screening $\alpha = 0.125$ and a weak ($\epsilon = 0.01$) as well as a 
strong ($\epsilon = 0.2$) second repulsion. In both cases, for small 
$\kappa$ the transition lines start at OCP and follow the single-Yukawa 
line. After a crossover regime, for $\kappa \gtrsim 15$ for $\epsilon = 
0.01$ and $\kappa \gtrsim 10$ for $\epsilon = 0.2$, the transitions are 
described well by rescaling the original single-Yukawa results according 
to 
\begin{equation}\label{eq:scale2Y}
\Gamma' = \Gamma/\epsilon,\quad\kappa' = \kappa/\alpha\,.
\end{equation}
In Fig.~\ref{fig:PD2Ys}, scaling by Eq.~(\ref{eq:scale2Y}) is demonstrated 
by transforming the MCT results for $\epsilon = 0$ (full circles) into 
a rescaled version (open circles) for $\epsilon = 0.2$ which compares well
to the full MCT calculation for the double Yukawa potential (diamonds). 
Similarly, formula~(\ref{eq:scale}) can be used to describe all double 
Yukawa results for small screening lengths, and the results for large 
screening lengths by scaling Eq.~(\ref{eq:scale}) with 
Eq.~(\ref{eq:scale2Y}). The dotted and dashed curves in 
Fig.~\ref{fig:PD2Ys} exhibit the scaled curves for $\epsilon = 0.01$ and 
0.2, respectively. The linear combination of the analytical descriptions 
for both length scales reads 
\begin{equation}\label{eq:scaleall}
\begin{array}{l}
\Gamma^c(\kappa)/\Gamma^c_\text{OCP} =
\left[e^{-\kappa}(1+\kappa+\kappa^2/2)\right.\\\left.
\qquad\qquad\qquad
+\epsilon\, e^{-\kappa\alpha}(1+\kappa\alpha+\kappa^2\alpha^2/2)
\right]^{-1}\,,\end{array}
\end{equation}
and is demonstrated by the solid line for $\epsilon = 0.01$ in 
Fig.~\ref{fig:PD2Ys}. It is seen that Eq.~(\ref{eq:scaleall}) describes 
the MCT results for double Yukawa potentials for the entire range of 
control parameters including crossover regions. In conclusion, the 
MCT predictions for both single and double Yukawa potentials can be 
rationalized by a single analytical formula (\ref{eq:scale}) which traces 
the melting curve, captures the interplay between large and small 
repulsive length scales, and extends for all parameters from OCP to HSS.

\subsection{Localization Lengths}

Another length scale resulting from the dynamical MCT calculations is 
given by the localization length. It is defined from the long-time limit 
of the mean-squared displacement $\delta r^2(t) = \langle|r(t) - r(0)|^2 
\rangle$ as ${r_{s}}^c=\sqrt{\lim_{t\rightarrow\infty}\delta r^2(t)/6}$. 
For the glass transition in the HSS, MCT predicts a localization length 
within HNC of $r_{s}^c/d = 0.0634$. This scale is quite close to the 
classical result of a Lindemann length \cite{Lindemann1910}.

\begin{figure}[htb]
\includegraphics[width=\columnwidth]{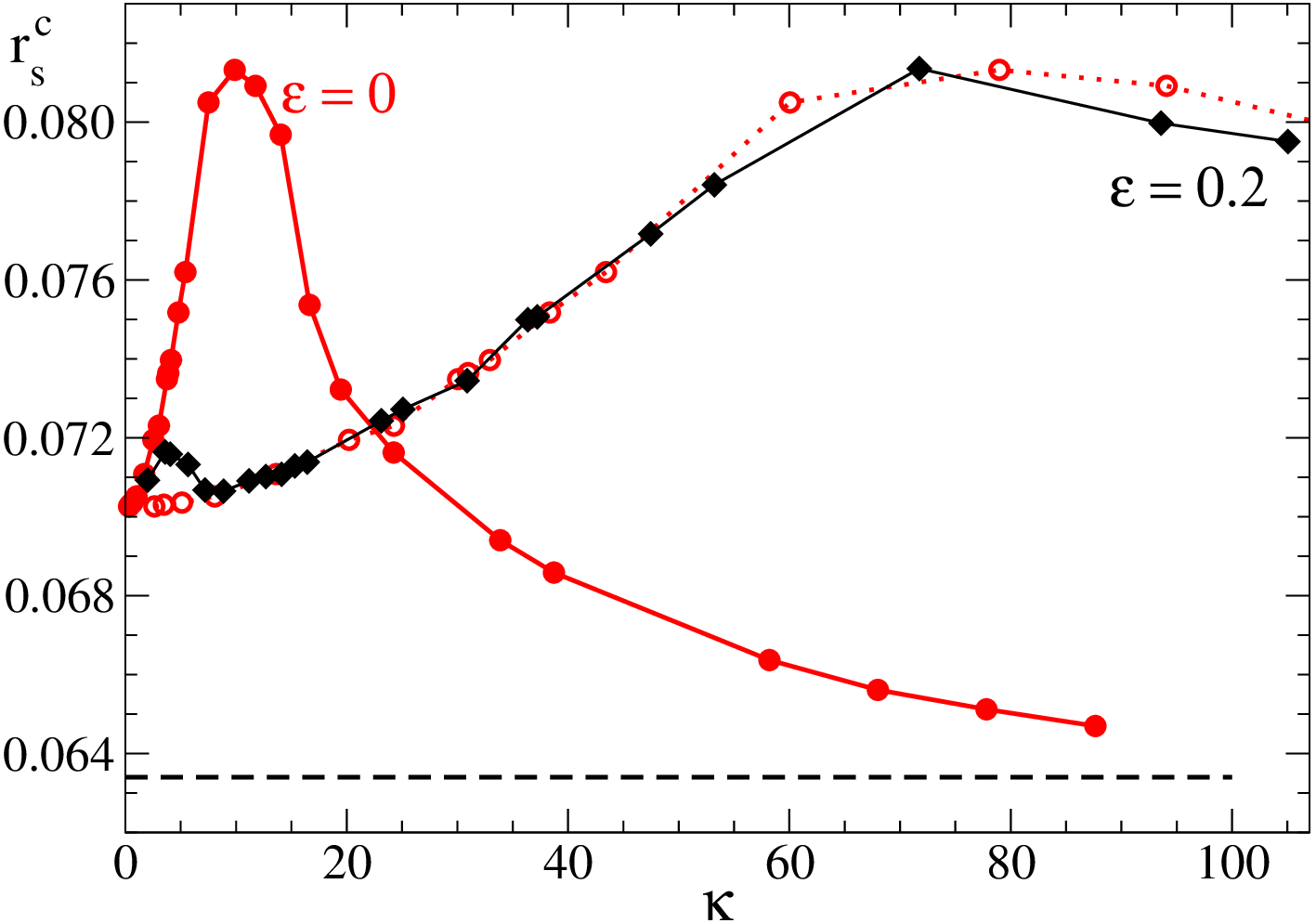}
\caption{\label{fig:loc} Localization length for single Yukawa (full 
circles) and double Yukawa (diamonds) potential with $\alpha = 0.125$ and 
$\epsilon = 0.2$. The open circles show the single-Yukawa data scaled 
according to Eq.~(\ref{eq:scale2Y}). The horizontal dashed line shows the 
HSS limit for $r_{s}^c$. 
} 
\end{figure}

For the single and double Yukawa potential, the evolution of the 
localization length with $\kappa$ is demonstrated in Fig.~\ref{fig:loc}. 
From a value of $r_{s}^c = 0.070$ for OCP, the localization lengths 
increase for the single Yukawa potential, reach a maximum around $\kappa 
\approx 10$ and decrease to the values for HSS for large $\kappa$. The 
maximum can be interpreted as follows: The widths of the distributions in 
$f_q$ seen in Fig.~\ref{fig:fq1Y} correspond to an inverse length scale 
equivalent to ${r_{s}}^c$, and the smaller width of the $f_q$ mean an 
increase of ${r_{s}}^c$. For large $\kappa$, the localization length needs 
to approach the HSS value, hence the ${r_{s}}^c$ decrease again. Both 
trends together yield a maximum.

The localization lengths for the double Yukawa system follows the 
single-Yukawa results for small $\kappa\lesssim 5$ as observed in 
Fig.~\ref{fig:PD2Ys} and hence increases; for $\kappa\gtrsim 5$, the 
double Yukawa system approaches the scaled single-Yukawa results shown by 
the circles. For larger $\kappa$, the evolution follows the scaled 
single-Yukawa results and while deviating for $\kappa\gtrsim 50$ from the 
scaled results, a scaled maximum is reached around $\kappa\approx 80$. 

Altogether, the variation of the localization lengths is around 10\% which 
is small compared to other glass-transition diagrams \cite{Sperl2004}. 
Hence we conclude that for both single- and double-Yukawa potentials the 
MCT results for the localization length are always close to the values 
usually assumed for the Lindemann criterion.

\section{Conclusion}

In summary, we have demonstrated above the full glass-transition diagram 
for the single and double Yukawa systems. While some parallel running 
lines for limited parameter ranges have been shown earlier for logarithmic 
core potential plus Yukawa tail \cite{Foffi2003star,Sciortino2004}, here 
we describe the transition diagrams by analytical formulae. In particular 
it could be shown how the HSS limit continuously evolves into the OCP 
limit. We have shown that the glass-transition lines resulting from the 
combination of HNC and MCT -- two rather complex nonlinear functionals -- 
can be described analytically over their entire range from the OCP limit 
for small $\kappa$ to the HSS limit for large $\kappa$. Qualitatively, the 
behavior of the transition line can be estimated by the Lindemann 
criterion for melting \cite{Lindemann1910}, while quantitatively, glass 
transition and crystal melting are following remarkably similar trends for 
stronger coupling.

It is important to note that the present calculations were performed for 
point particles with various degrees of charging and screening. The limit 
of the HSS emerges from that calculations without actual excluded volume 
in the potentials. With the important difference of a finite hard-sphere 
radius being present, the possibility that in addition to a Coulomb 
crystal a dilute system of charges may also form a Coulomb glass was 
explored in the restricted primitive model for a mixture of charged hard 
spheres \cite{Bosse1998} and the hard-sphere jellium model 
\cite{Wilke1999} as well as for a system of charged hard spheres to 
describe charge-stabilized colloidal suspensions \cite{Lai1995}. In 
conclusion, the present calculations offer exhaustive analytical 
descriptions for glass transitions over a wide range of quite different 
interaction potentials. The predictions should motivate data collapse from 
computer simulation and different experimental model systems in order to 
confirm or challenge the unified picture presented above.

Financial support within the ERC (Advanced grant INTERCOCOS, project 
number 267499) is gratefully acknowledged.

\bibliographystyle{apsrev}

\end{document}